\documentclass[aps,prd,showpacs,superscriptaddress,amsmath,amssymb,eqsecnum,nofootinbib]{revtex4-2}

\usepackage{graphicx}
\usepackage{dcolumn}
\usepackage{bm}

\usepackage{color}

\newcommand{\bea}{\begin{eqnarray}}
\newcommand{\eea}{\end{eqnarray}}

\newcommand{\la}{\lambda}
\newcommand{\pa}{\partial}

\newcommand{\hide}[1]{}

\begin{document}

\title{
Running Newton Coupling, Scale Identification and Black Hole Thermodynamics}

\author{Chiang-Mei Chen} \email{cmchen@phy.ncu.edu.tw}
\affiliation{Department of Physics, National Central University, Zhongli, Taoyuan 320317, Taiwan}
\affiliation{Center for High Energy and High Field Physics (CHiP), National Central University,
 Zhongli, Taoyuan 320317, Taiwan}

\author{Yi Chen} \email{yi592401@gmail.com}
\affiliation{Department of Physics, National Central University, Zhongli, Taoyuan 320317, Taiwan}

\author{Akihiro Ishibashi} \email{akihiro@phys.kindai.ac.jp}
\affiliation{Department of Physics, Kindai University, Higashi-Osaka, Osaka 577-8502, Japan}
\affiliation{Research Institute for Science and Technology, Kindai University, Higashi-Osaka, Osaka 577-8502, Japan}

\author{Nobuyoshi Ohta} \email{ohtan@ncu.edu.tw}
\affiliation{Department of Physics, National Central University, Zhongli, Taoyuan 320317, Taiwan}
\affiliation{Research Institute for Science and Technology, Kindai University, Higashi-Osaka, Osaka 577-8502, Japan}

\author{Daiki Yamaguchi} \email{daichanqg@gmail.com}
\affiliation{Department of Physics, Kindai University, Higashi-Osaka, Osaka 577-8502, Japan}


\begin{abstract}

We discuss the quantum improvement of black hole solutions in the context of asymptotic safety.
The Newton coupling in this formulation depends on an energy scale, which must be identified with
some length scale in order to study physical consequences to black holes. However, no physical principle
has so far been known for the identification.
Here we propose that the consistency of the first law of thermodynamics is the principle
that should determine physically sensible scale identification, at least close to the horizon.
We show that this leads to a natural solution that the Newton coupling should be a function of
the horizon area and find a universal formula for the quantum entropy,
which agrees with the standard Bekenstein-Hawking entropy for constant Newton coupling,
for Kerr black holes and other higher-dimensional black holes.
This suggests that the Newton coupling is a function of the area near the horizon,
and also away to infinity, where the quantum effects may not be so important.

\end{abstract}


\maketitle

\section{Introduction}

The black hole spacetimes are important solutions of Einstein's equations.
Understanding the dynamics of these spacetimes when quantum effects of the geometry are taken into account
is  one of the most challenging issues in theoretical physics.
It is known that these solutions have singularities close to their center, where Einstein's equations
are no longer valid. Indeed it is plausible that quantum effects play a significant role there.

The standard perturbative quantization of Einstein gravity is of little help because it is a nonrenormalizable theory.
Among others, one of the interesting approaches to quantum gravity is the one using the functional renormalization
group (FRG), called asymptotic safety when the quantum field theories including gravity in this framework
exhibit well-defined ultraviolet (UV) behaviors~\cite{Reuter:1996, Souma}.
For recent reviews of the subject, see Refs.~\cite{ASrev1, ASrev2, ASrev3}.
A central object in the FRG is the effective average action $\Gamma_k$.
It is defined in the path integral by integrating over UV modes beyond a cutoff energy scale $k$,
and when evaluated at tree level, it correctly describes gravitational phenomena with quantum effects included,
whose typical momenta are of order $k$. An important consequence of the quantum effects is that
the Newton coupling depends on the energy scale $k$.
Actually the energy scale $k$ is introduced just as a cutoff scale and may not really correspond to
the energy scale that physical phenomena are observed~\cite{Donoghue:2019}. We do not intend to argue that it is
such an energy scale, but would like to find its any possible connection to some length scale so that we can get
some insight into how the quantum effect manifests itself in the black hole spacetime.

There are several ways to incorporate the quantum effects into black holes. Here we choose the way to make
renormalization group (RG) improvement at the level of the solutions, in which case we replace
the couplings appearing in the classical solutions with the running couplings.
By suitably choosing the identification of the energy scale with some length scale in the solutions,
applications of this to black hole systems have been considered
in~\cite{BR:2000,BR:2006,FLR12,RT:2010,HT:2011,FL:2012,KS141,LN,KS142,BKP2017,PS:2018,Plat2020,IOY:2021,RT:2021,RW04}.
A feasible approach to account for spacetime features resulting from the quantum improvement is to use
diffeomorphism-invariant proper distance integrals for the identification of the energy scale.
It has been shown that the quantum effects are negligible for large masses, but it has a significant consequence
that the central singularities become milder, if not resolved, for the spherically symmetric static black holes,
i.e., Schwarzschild solution~\cite{BR:2000} and Reissner-Nordstr\"om black hole~\cite{IOY:2021}.

However, when the rotating black hole, Kerr solution, is considered, it is found that the first law of
black hole thermodynamics is not satisfied straightforwardly.
Since the first law of thermodynamics is the fundamental law of energy conservation, it must be valid
not only semiclassically but also quantum mechanically.
In Ref.~\cite{RT:2010}, the energy $k$ is identified with the inverse radial length from the origin to
the point under consideration with {\it a fixed angle}.
With this scale identification, it turned out that either there exists no entropy-like state function or
the temperature cannot be in proportion to the surface gravity.
For this reason, it has been proposed in~\cite{RT:2010} that the definitions of the temperature and entropy
should be modified in such a way that the first law is satisfied in the small angular-momentum expansions.
However, it is unnatural to consider that the temperature is modified, since it should be defined
through the geometric quantity of the surface gravity at the horizon.

In Ref.~\cite{PS:2018}, another identification using the Kretschmann scalar $K = R_{\mu\nu\rho\la} R^{\mu\nu\rho\la}$,
a diffeomorphism-invariant quantity of momentum dimension four, was proposed for Kerr solution.
However, because this quantity also depends on the angles, they took a special value of the angle.
Unfortunately this again leads to the difficulty in satisfying the first law of thermodynamics.
Because this is a fundamental physical principle, it must be satisfied.
We believe that it is unnatural to choose such a specific angle because there is no physical reason why any angle
should be chosen for the identification. Is it then better to keep the angular dependence?
However as we show explicitly shortly, such angular dependence in general leads to singularities
on the horizon, and cannot be accepted.
Various diffeomorphism-invariant quantities are considered in~\cite{EH:2021}, but all of them depend on
the angles for the Kerr spacetime. It is an important question if and why we have to choose any specific angle.
The physical features of spacetime such as the number of horizons, Hawking temperatures and the strength
of the curvature singularity actually do depend on the particular choice of $k(r)$, so this identification
has a critical importance. If the identification is made with a specific angle, that would cause various
physical quantities to also depend on the angles, which may not be accepted.

The fundamental problem in these approaches is the lack of the physical guiding principle that must be satisfied
in making the identification.
In this paper, we propose that the first law of thermodynamics is the physical principle that should determine
the identification, at least in the close neighborhood of the event (outer) horizon. To clarify the implications
of this proposal, we first assume that the Newton coupling becomes a function of the radial coordinate and
angular momenta, if the solution involves angular momenta, and examine what restriction on the form of
the Newton coupling is imposed by the consistency of the first law of thermodynamics.
We find that the simplest solution is that the Newton coupling should be a function of the area of the horizon.
We study various black hole solutions including higher-dimensional ones, and find that
this result is universal and valid for all the solutions we have examined.
The area is a geometric quantity which avoids the above-mentioned problems in the identification involving angles.
Moreover, we find that this gives a universal formula for the entropy:
$$
S = \int \frac{dx_+}{4 G(x_+)},
$$
where $x_+$ is the horizon area.
The same formula was also given in~\cite{FL:2012} for Kerr black holes taking the postulate that the energy
scale $k^2$ is determined by the horizon area, and in particular it is inversely proportional to the horizon area;
they considered only the possible dependence on the horizon area from the outset, and no other dependence
such as on horizon radius was examined.
Here without such a postulate, we show that we are almost uniquely
led to this formula starting from the consistency of the first law of thermodynamics, and the formula
is valid not only for the Kerr black holes but also for all the black holes we consider.
This reproduces the standard Bekenstein-Hawking entropy~\cite{Bek,Haw} when the Newton coupling is constant.
We regard this as a strong evidence that our result is correct, even though it is
the simplest solution of the consistency condition. Strictly speaking, this result is valid only at the horizon,
but considering its geometrical nature, it is natural to extend it away from the horizon.
This means that the Newton coupling would be a function of the area at the corresponding radius.
This identification seems to be on the right track outside the horizon because the asymptotic behavior agrees
with the classical case, but we find that the high energy limit is not achieved for all region of space
even near the singularity for the Kerr solution. We thus expect that the identification would be modified
near the singularity.

This paper is organized as follows.
In sect.~\ref{runningcouplings}, we summarize the running couplings we use for the black hole solutions.
In sect.~\ref{kerr}, we study the (A)dS-Kerr black holes and find the first evidence that the solution
to the consistency condition of the first law of thermodynamics is that the identification should be
made with the area. We can reduce this to the simple Schwarzschild black hole by
just taking the limit of vanishing angular momentum, though there is no consistency condition for this
case since the entropy only depends on the radius. We do not expect that this result would change even if
we include charge. Thus this result is valid for all black hole solutions in four dimensions.
In sect.~\ref{MPBH}, we consider five-dimensional Myers-Perry black holes with two angular momenta~\cite{MP},
and again find that the consistency requires that the identification should be made through the area,
and the entropy is given by the above universal formula.
Here we consider the solutions with two general angular momenta, and those with the same angular momenta.
In sect.~\ref{KKBH}, we find further evidence for Kaluza-Klein black string~\cite{Kastor:2006ti, Harmark:2007md}.
Section~\ref{DC} is devoted to some discussions and conclusions.

\section{Running Couplings}
\label{runningcouplings}

We start with the action for the theory we consider in this paper.
The action of the Einstein theory with a cosmological constant, in the units $c = \hbar = 1$, is
\begin{equation}
S = - \frac1{16 \pi G} \int d^4x \sqrt{-g} \left( R - 2 \Lambda \right),
\end{equation}
where $G$ is the Newton coupling and $\Lambda$ the cosmological constant.

Due to the quantum effects, the couplings $G$ and $\Lambda$ are functions of the energy scale $k$, given
by the renormalization group (RG) equations~\cite{Reuter:1996,Souma,PS:2018, IOY:2021}.
For simplicity, we assume that the cosmological constant is already at the fixed point and negligibly small.
The RG equation for dimensionless gravitational coupling $\tilde{G}(k) = k^2 G(k)$ is then
\begin{equation}
k \partial_k \tilde{G}(k) = 2 \tilde{G}(k) + \frac{B_1 \tilde{G}^2(k)}{1 + B_2 \tilde{G}(k)}
= 2 \tilde{G}(k) \frac{1 + \omega' \tilde{G}(k)}{1 + (\omega + \omega') \tilde{G}(k)}, \qquad
\omega' = \frac12 B_1 + B_2, \quad \omega = - \frac12 B_1,
\end{equation}
where $B_1$ and $B_2$ are constants that can be determined once the theory is given.
In our present case, they are $B_1=-\frac{8}{\pi}(1-\frac{\pi^2}{144})$ and $B_2=\frac{2}{3\pi}$.
This leads to the implicit solution
\begin{equation}
\tilde{G}(k) [ 1 + \omega' \tilde{G}(k) ]^{\omega/\omega'}
= \tilde{G}(k_0) [ 1 + \omega' \tilde{G}(k_0) ]^{\omega/\omega'} (k/k_0)^2,
\end{equation}
where $k_0$ is a reference energy scale.
In order to obtain an analytic expression for the ease of the following discussions,
we make the approximation that $\omega' \approx - \omega$, i.e. $B_2 \approx 0$.
Then we obtain the analytic solution
\begin{equation}
\tilde{G}(k) = \frac{\tilde{G}(k_0) (k/k_0)^2}{1 - \omega \tilde{G}(k_0) + \omega \tilde{G}(k_0) (k/k_0)^2}
\quad
\Rightarrow
\quad G(k) = \frac{G(k_0)}{1 - \omega G(k_0) k_0^2 + \omega G(k_0) k^2}.
\end{equation}
We choose $k_0 = 0$ and define $G_0 = G(k_0)$ to get
\begin{equation}
\label{Gk}
G(k) = \frac{G_0}{1 + \omega G_0 k^2}.
\end{equation}
Generalization to higher dimensions is possible with suitable powers of the energy scale $k$.

\section{Kerr-(A)dS Black Holes}
\label{kerr}

Let us consider our first example of (A)dS Kerr solutions. The line-element of the classical Kerr-(A)dS black
hole solution in the Boyer-Linquist coordinate is
\begin{eqnarray}
ds^2 = - \frac{\Delta_r}{\Sigma} \left( dt - \frac{a \sin^2\theta}{\Xi} d\varphi \right)^2
+ \frac{\Sigma}{\Delta_r} dr^2 + \frac{\Sigma}{\Delta_\theta} d\theta^2
+ \frac{\Delta_\theta}{\Sigma} \sin^2\theta \left( a dt - \frac{r^2 + a^2}{\Xi} d\varphi \right)^2,
\eea
where
\bea
&& \Delta_r = (r^2 + a^2) \left( 1 - \frac{\Lambda}3 r^2 \right) - 2 G M r, \qquad
\Delta_\theta = 1 + \frac{\Lambda}3 a^2 \cos^2\theta,
\nonumber\\
&& \Sigma = r^2 + a^2 \cos^2\theta, \qquad
\Xi = 1 + \frac{\Lambda}3 a^2.
\label{Kerrbh}
\end{eqnarray}
Here $a$ is the angular momentum parameter.

In the semiclassical case, where $G$ is a constant, the Hawking temperature and the entropy are given by
\begin{equation}
T = \frac{\kappa}{2\pi} = \frac{r_+}{4 \pi (r_+^2 + a^2)} \left( 1 - \frac{a^2}{r_+^2}
- \frac{\Lambda a^2}3 - \Lambda r_+^2 \right), \qquad
S = \frac{\mathcal{A}}{4 G} = \frac{\pi (r_+^2 + a^2)}{G \Xi}.
\label{ts}
\end{equation}
The quantity $\kappa$ here is the surface gravity, which is defined as follows:
We choose $\Omega_H$ in the Killing vector $\chi = \partial_t + \Omega_H \partial_\varphi$ such that
$\chi$ is null ($\chi_\mu \chi^\mu = 0$) at the horizon, and then the surface gravity $\kappa$ is given by
\bea
\kappa = \sqrt{-\frac{(\nabla_\mu \chi_\nu) (\nabla^\mu \chi^\nu)}{2}}.
\label{surfaceg}
\eea
Also $r_+$ is the outer horizon radius defined by $\Delta_r(r_+)=0$.
Note the second expression in \eqref{ts} for the entropy is the standard Bekenstein-Hawking formula~\cite{Bek,Haw}.
These satisfy the first law of thermodynamics
\begin{equation}\label{firstlaw}
dE = T dS + \Omega dJ,
\end{equation}
where
\begin{equation}
E = \frac{M}{\Xi^2}, \qquad J = \frac{M a}{\Xi^2}, \qquad \Omega = \frac{a (1 - \Lambda r_+^2/3)}{r_+^2 + a^2}.
\end{equation}
We also have the Kretschmann scalar
\begin{equation}
K = \frac83 \Lambda^2 + \frac{48 M^2 G^2 (r^6 - 15 a^2 \cos^2\theta r^4 + 15 a^4 \cos^4\theta r^2 - a^6 \cos^6\theta)}
{(r^2 + a^2 \cos^2\theta)^6},
\label{KerrKretschmann}
\end{equation}
which does depend on the angle $\theta$.

For the quantum improved black hole, we replace the Newton coupling by the length-scale dependent one
with a proper identification with the energy scale.
Since the Kretschmann scalar in \eqref{KerrKretschmann} depends on the angle, as we will show shortly, it is
not suitable for the identification even though it is a diffeomorphism-invariant quantity of momentum dimension four.
Both Ref.~\cite{BR:2000} and~\cite{PS:2018} chose quantities for fixed angle $\theta$ in the scale identification.
In order to satisfy the first law of thermodynamics, then they have to modify
the expressions for the temperature and the entropy. However, the temperature then loses its geometrical
meaning given by the surface gravity, and we suppose that this is unnatural.
So here we assume that the formula $T = \kappa/2\pi$ is still valid even in the quantum improved black holes,
and the running couplings are dependent on the mass parameter $M$.
A more convenient choice is, instead of $M$, the (outer) horizon radius $r_+$ since one can straightforwardly
obtain $M(r_+, a)$ from $\Delta_r(r_+) = 0$:
\bea
M(r_+, a)=\frac{r_+^2+a^2}{2G r_+}\left(1-\frac{\Lambda}{3} r_+^2 \right).
\label{mass}
\eea
From the first law, the variation of the entropy can then be obtained as
\begin{equation}
dS = \partial_+ S \, dr_+ + \partial_a S \, da,
\end{equation}
where $\partial_+$ denotes the derivative with respect to $r_+$.
Using similar expressions for $dM$ and $dJ$ in the first law, we find
\begin{equation}
\partial_+ S = \frac{\partial_+ M - \Omega \, \partial_+ J}{T}, \qquad
\partial_a S = \frac{\partial_a M - \Omega \, \partial_a J}{T}.
\label{Kerrent}
\end{equation}
They must satisfy the consistency condition
\begin{equation}
\label{cond}
\partial_a \left( \partial_+ S \right) = \partial_+ \left( \partial_a S \right).
\end{equation}
The relation of energy scale $k$ to the spacetime coordinates is an essential issue
for constructing quantum solutions. We consider constraint on such a scale identification from this condition.

\subsection{Kerr Black Holes}

Let us first consider the simple Kerr black holes for $\Lambda = 0$ which are axisymmetric,
so the coupling $G$ in general can be a function of $r$ and $\theta$.
Using the metrics given in Eq.~\eqref{Kerrbh} with $G=G(r,\theta)$, we find that the scalar curvature
is given by
\begin{equation}
R = \frac{2 M}{\Sigma} \left[ r \partial_r^2 G(r, \theta) + 2 \partial_r G(r, \theta)
- \frac{M r^2}{\Delta_r^2} \Bigl( \partial_\theta G(r, \theta) \Bigr)^2 \right].
\label{Kerrsc}
\end{equation}
This means that we have the curvature singularity at the horizon. To avoid this,
the $G(r, \theta)$ should have the following form:
\begin{equation}
\label{Grtheta}
G(r, \theta) = G_1(r) + \Delta_r^2(r, \theta) G_2(r, \theta).
\end{equation}
We see that the $\theta$-derivative of this vanishes at the horizon, so that
the scalar curvature~\eqref{Kerrsc} is finite at the horizon, where $\Delta_r(r,\theta)=0$.
This relation contains $G(r,\theta)$ on the right-hand side in $\Delta_r(r,\theta)$.
If needed, we should solve this equation by explicitly writing it in terms of $G(r,\theta$),
but we do not have to do so since we will restrict it to the horizon, where $\Delta_r=0$.

Since thermodynamics is considered in terms of the quantities on the horizon, Eq.~\eqref{Grtheta}
would mean that $G$ is a function of only $r_+$ when we discuss thermodynamics.
Suppose that it does not depend on other parameters such as $a$.
To obtain the Bekenstein-Hawking temperature, we first calculate the surface gravity~\eqref{surfaceg}
for $r$-dependent $G$. Then the temperature is found to be
\begin{equation}
\label{def:temp:Kerr}
T = \frac{\kappa}{2\pi}=\frac{(r_+^2 - a^2) G(r_+)- r_+ (r_+^2 + a^2) \partial_+ G(r_+)}{4 \pi r_+ (r_+^2 + a^2) G(r_+)}.
\end{equation}
Note that for the constant $G$, the temperature evaluated in this manner reduces to the standard
Hawking temperature for the classical counterpart.
Then, the consistency condition~(\ref{cond}) gives
\bea
r_+ (r_+^2 - a^2) \partial_+^2 G(r_+) - 2 a^2 \partial_+ G(r_+) = 0,
\eea
leading to
\bea
G(r_+) = c_1 \frac{r_+^2 + a^2}{r_+} + c_2. 
\eea
Here $c_1$ should be zero since we have assumed $\partial_a G(r_+) = 0$. This implies that the coupling does not
run in order to ensure the first law. This problem had been observed in~\cite{RT:2010} and the authors claimed
that the temperature formula $T = \kappa/2 \pi$ should be improved. However, the temperature then loses
its geometrical interpretation.

Here we point out that there is an alternative possibility by allowing the angular-momentum dependence,
$G = G(r_+, a)$. The parameter $a$ in some sense is related to a choice of coordinates since the Kerr black hole
reduces, in the limit of $M = 0$, to Minkowski space with elliptical foliations characterized by $a$
(spherical foliations for $a = 0$). In such case, we find from Eq.~\eqref{Kerrent}
\begin{equation}
\partial_+ S = \frac{2 \pi r_+}{G(r_+, a)}, \qquad
\partial_a S = \frac{2 \pi \left[ r_+^2 (r_+^2 + a^2) \partial_a G(r_+, a) - a (r_+^2 - a^2) G(r_+, a) \right]}
{G(r_+, a) \left[ r_+ (r_+^2 + a^2) \partial_+ G(r_+, a) - (r_+^2 - a^2) G(r_+, a) \right]}.
\label{cons_Kerr}
\end{equation}
The consistency condition~(\ref{cond}) can be reexpressed as
\bea
\partial_+ \frac{r_+^2 (r_+ \partial_a G - a \partial_+ G)}{r_+ (r_+^2 + a^2) \partial_+ G
- (r_+^2 - a^2) G} = 0,
\label{co_Kerr}
\eea
which can be integrated as
\bea
\frac{r_+^2 (r_+ \partial_a G - a \partial_+ G)}{r_+ (r_+^2 + a^2) \partial_+ G - (r_+^2 - a^2) G} = f(a),
\eea
with some function $f(a)$ of the angular momentum parameter $a$.

Let us consider the simplest solution for $f(a) = 0$. Though this is the simplest solution, we will find
that this gives a very reasonable result on the entropy.
We tried to study some more general solutions, but it seems that such solutions do not give more interesting results.
We then find $G(r_+, a) = G(4\pi(r_+^2 + a^2)) = G(x_+)$, with $x_+ = 4 \pi (r_+^2 + a^2)$ which leads to
\begin{equation} \label{r2a2}
\partial_+ G(x_+) = 8 \pi r_+ G'(x_+), \qquad
\partial_a G(x_+) = 8 \pi a G'(x_+).
\end{equation}
Here the prime means the derivative with respect to its argument.
Note that $x_+$ is the horizon area.
Substituting this into Eqs.~\eqref{cons_Kerr}, we find
\begin{equation}
\partial_+ S = \frac{2 \pi r_+}{G(x_+)}, \qquad \partial_a S = \frac{2 \pi a}{G(x_+)}.
\end{equation}
Because the consistency condition is satisfied, both expressions lead to the same entropy formula
\begin{equation}
S = \int \frac{dx_+}{4 G(x_+)}, \qquad
x_+ \equiv 4 \pi (r_+^2 + a^2).
\end{equation}
If the Newton coupling is a constant, this gives the well-known Bekenstein-Hawking formula
for the entropy~\cite{Bek, Haw}, in accordance with the semiclassical limit. We find that this is a strong evidence
that our choice is on the right track. As a simple example, let us consider the following scale identification:
\bea
k_+ = \dfrac{\xi}{\sqrt{x_+}} = \frac{\tilde\xi}{\sqrt{r_+^2 + a^2}},
\label{esrelation}
\eea
on dimensional grounds, where $\xi(=\sqrt{4\pi} \tilde\xi)$ is some dimensionless constant. We then obtain
\bea
G(r_+, a) = \frac{G_0 (r_+^2 + a^2)}{r_+^2 + a^2 + \tilde\omega G_0},
\label{qNewton}
\eea
where
\bea
\tilde \omega = \omega \tilde\xi^2.
\eea
The entropy is then found to be
\begin{equation}
S = \frac{\pi (r_+^2 + a^2)}{G_0} + \pi \tilde\omega \ln (r_+^2 + a^2).
\end{equation}
The first term is the classic result of the area divided by $4G_0$, and
the logarithmic corrections are the terms obtained as quantum corrections~\cite{Kaul:2000kf}
and our result is again reasonable for the quantum case.
Similar result is also obtained in loop quantum gravity~\cite{Meissner:2004ju}.
We would like to emphasize that our above quantum
corrected entropy satisfies the desired first law of the form (\ref{firstlaw}),
without the need to modify the definition of the temperature, that is, the one given by the surface gravity.

The above relations are valid only at the horizon. However, considering the geometrical nature of the area,
it would be natural to extend the relation away from the horizon and make our scale identification
in terms only of the surface area $x$ at fixed radius $r$, at least, in a neighborhood of the (outer) horizon.
\begin{eqnarray}
x &=& 2\pi \int_0^\pi \sqrt{(r^2 + a^2)^2 - r_+ r_- (r - r_+) (r - r_-) \sin^2\theta} \, \sin\theta \, d\theta
\nonumber\\
&=& 2\pi(r^2 + a^2) - \pi\frac{(r^2 + a^2)^2 - a^2 (r - r_+)(r - r_-)}{a \sqrt{(r - r_+)(r - r_-)}}
\ln\frac{r^2 + a^2 - a \sqrt{(r - r_+)(r - r_-)}}{r^2 + a^2 + a \sqrt{(r - r_+)(r - r_-)}},
\label{area at r}
\end{eqnarray}
where we have used the relations~\eqref{mass} and $r_-=a^2/r_+$ (inner horizon radius).
This has the following properties:
\begin{enumerate}
\item[(i)]
$x$ is the area of the fixed radius $r$, and this reproduces the horizon area $x_+$ in the limit of $r\to r_+$.

\item[(ii)]
It gives real value for all $r > 0$ including $r_- < r < r_+$ where square roots are imaginary.

\item[(iii)]
For large $r$ outside the horizon, $x\to r^2$ and we see that this behavior gives the right
classical behavior (constant $G(r,a)$ in the long distance).

\item[(iv)]
For $r \to 0$, $x$ obtained by integrating over the interior disk of ring singularity,
gives nonvanishing value $a^2$. So $x$ never approaches zero anywhere in physical region $r>0$.
\end{enumerate}

A nice feature of the identification is that, when $r$ becomes large  where we expect that the quantum effects
would not be large, the Newton coupling~\eqref{qNewton} goes to a constant, reducing to the classical behavior.
However we find there is a subtlety if we extend the identification this way.
For the computation of the temperature~\eqref{def:temp:Kerr}, we should first substituted $r_+$ into $G(r)$ and
afterwards $\partial_+$ should be taken.
However when the extension by \eqref{area at r} is considered, it is more natural to first calculate
the surface gravity and then take the limit $r\to r_+$.
This gives a different temperature from Eq.~\eqref{def:temp:Kerr} because
\bea
\lim_{r \to r_+} \frac{\pa x}{\pa r} \frac{\pa G(x)}{\pa x}\ \neq\ \frac{\pa x_+}{\pa r_+} \frac{\pa G(x_+)}{\pa x_+},
\eea
as can be easily checked with \eqref{area at r}. This would lead to the breakdown of the consistency.

It may be then more natural and simpler to use the area of some surface obtained by just replacing $r_+$ by $r$:
\bea
x=4\pi (r^2+a^2).
\label{area:Kerr}
\eea
After all, the above extension of the area away from the horizon is dependent on the choice of the Boyer-Linquist
coordinate. If we use different coordinate system, ``fixed $r$'' would not make much sense.
If we adopt this definition of the area~\eqref{area:Kerr}, the problem mentioned above goes away.
Further comments are given in sect.~\ref{DC}.

On the other hand, as we can see from the Kretschmann invariant~\eqref{KerrKretschmann}, there is a ring singularity
in the Kerr black hole at $\theta={\pi}/{2}$ and $r=0$. If we simply adopt the identification~\eqref{esrelation}
with $x_+$ replaced by $x$ whichever the choice of $x$, this is true also for quantum improved cases;
our quantum improved Newton coupling does not vanish there, and the singularity is not resolved.
Also Eq.~\eqref{esrelation}, extended away from the horizon, would mean that the high-energy limit $k\to\infty$ would
not be achieved for any range of spacetime since $x$ never vanishes. However, such a high-energy limit should exist
according to the FRG.
Thus we expect that the identification may be correct near and away from the horizon (larger $r$),
but would be significantly modified near the singularity.
Similar change in the behavior is observed in \cite{BR:2000}.

\subsection{Kerr-(A)dS Black Holes}

For the Kerr-(A)dS black holes (with nonzero constant $\Lambda$), if we assume $G = G(r_+, a)$,
the temperature calculated again from the surface gravity $\kappa$ in Eq.~\eqref{surfaceg} is
\begin{equation}
\label{def:temp:KAdS}
T = \frac{\kappa}{2\pi} =\frac{\left( r_+^2 - a^2 - r_+^4 \Lambda - a^2 r_+^2 \Lambda/3 \right) G(r_+, a)
- r_+ (r_+^2 + a^2) \left( 1 - r_+^2 \Lambda/3 \right) \partial_+ G(r_+, a)}{4 \pi r_+ (r_+^2 + a^2) G(r_+, a)}.
\end{equation}
From the first law and Eq.~\eqref{Kerrent}, we then find
\begin{eqnarray}
\partial_+ S &=& \frac{2 \pi r_+}{\Xi G(r_+, a)},
\nonumber\\
\partial_a S &=& \frac{2 \pi \left( 1 - r_+^2 \Lambda/3 \right)
\left[ a \left( r_+^2 - a^2 - r_+^4 \Lambda - a^2 r_+^2 \Lambda/3 \right) G(r_+, a)
- \Xi r_+^2 (r_+^2 + a^2) \partial_a G(r_+, a) \right]}{\Xi^2 G(r_+, a)
\left[ \left( r_+^2 - a^2 - r_+^4 \Lambda - a^2 r_+^2 \Lambda/3 \right) G(r_+, a)
- r_+ (r_+^2 + a^2) \left( 1 - r_+^2 \Lambda/3 \right) \partial_+ G(r_+, a) \right]}.
\label{ds_Kerr}
\end{eqnarray}
The consistency condition~\eqref{cond} can be reexpressed as
\bea
\partial_+ \frac{r_+^2 \left[ \Xi r_+ \partial_a G(r_+, a) - a \left( 1 - r_+^2 \Lambda/3 \right)
\partial_{r_+} G(r_+, a) \right]}{ \left( r_+^2 - a^2 - r_+^4 \Lambda - a^2 r_+^2 \Lambda/3 \right) G(r_+, a)
- r_+ (r_+^2 + a^2) \left( 1 - r_+^2 \Lambda/3 \right) \partial_{r_+} G(r_+, a)} = 0,
\eea
which again can be integrated as
\bea
\frac{r_+^2 \left[ \Xi r_+ \partial_a G(r_+, a) - a \left( 1 - r_+^2 \Lambda/3 \right) \partial_{r_+} G(r_+, a)
\right]}{ \left( r_+^2 - a^2 - r_+^4 \Lambda - a^2 r_+^2 \Lambda/3 \right) G(r_+, a)
- r_+ (r_+^2 + a^2) \left( 1 - r_+^2 \Lambda/3 \right) \partial_{r_+} G(r_+, a)} = f(a).
\eea
For the simplest choice $f(a) = 0$, we find $G(r_+, a) = G\left( 4 \pi \frac{r_+^2 + a^2}{\Xi} \right)$,
which simplifies \eqref{ds_Kerr} to
\begin{equation}
\partial_a S = \frac{2 \pi a \left( 1 - r_+^2 \Lambda/3 \right)}{\Xi^2 G(r_+, a)}.
\end{equation}
The resulting entropy is given by
\begin{equation}
S = \int \frac{dx_+}{4 G(x_+)}, \qquad
x_+ \equiv 4 \pi \frac{r_+^2 + a^2}{\Xi}.
\end{equation}
Note that $x_+$ here is the area of the horizon.
Taking essentially same identification as \eqref{esrelation}:
\bea
k_+ = \dfrac{\xi}{\sqrt{x_+}} = \frac{\tilde\xi \sqrt{\Xi}}{\sqrt{r_+^2 + a^2}},
\eea
with a dimensionless constant $\xi (=\sqrt{4\pi}\tilde\xi)$, we get
\bea
G(r_+, a) = \frac{G_0 (r_+^2 + a^2)}{r_+^2 + a^2 + \tilde{\omega} \Xi G_0},
\eea
where $\tilde{\omega}\equiv \omega \tilde\xi^2$.
The entropy is given by
\begin{equation}
S = \frac{\pi (r_+^2 + a^2)}{G_0 \Xi} + \pi \tilde{\omega} \ln \frac{r_+^2 + a^2}{\Xi}.
\end{equation}
Again the first term is the well-known classic Bekenstein-Hawking formula, and the second term
represents the quantum correction.

We also note that in the limit $a\to 0$, all of our results give those for the (A)dS-Schwarzschild black holes.
In such a spherically symmetric case, the consistency requirement from the thermodynamics becomes trivial
since there is no quantity other than the area radius on which the Newton coupling depends,
and we can determine the entropy from the first law of thermodynamics.
The result that it depends on the area is a universal one also valid in such cases.

\section{Five-dimensional Myers-Perry black holes}
\label{MPBH}

As a further nontrivial example, we consider higher-dimensional black holes.
The metrics for the 5D Myers-Perry black holes with two angular momenta are given by~\cite{MP}
\begin{eqnarray}
ds^2 &=& - dt^2 + \frac{G m r^2}{\Pi F} (dt - a_1 \sin^2\theta' d\phi_1 - a_2 \cos^2\theta' d\phi_2)^2
 + \frac{\Pi F}{\Pi - G m r^2} dr^2
\nonumber\\
&+& (r^2 + a_1^2) (\cos^2\theta' d\theta'^2 + \sin^2\theta' d\phi_1^2)
 + (r^2 + a_2^2) (\sin^2\theta' d\theta'^2 + \cos^2\theta' d\phi_2^2),
\end{eqnarray}
where
\begin{equation}
\Pi = (r^2 + a_1^2) (r^2 + a_2^2), \qquad
F = 1 - \frac{a_1^2 \sin^2\theta'}{r^2 + a_1^2} - \frac{a_2^2 \cos^2\theta'}{r^2 + a_2^2}.
\end{equation}
The Kretschman scalar is given by
\begin{equation}
K = R_{\alpha\beta\mu\nu} R^{\alpha\beta\mu\nu}
= \frac{24 G^2 m^2 (3 r^2 - a_1^2 \cos^2\theta - a_2^2 \sin^2\theta)(r^2 - 3 a_1^2 \cos^2\theta
 - 3 a_2^2 \sin^2\theta)}{(r^2 + a_1^2 \cos^2\theta + a_2^2 \sin^2\theta)^6}.
\end{equation}
The horizon is given by
\begin{equation}
\Pi(r_+) - G m r_+^2 = 0,
\end{equation}
and the mass, angular momenta and angular velocities are
\begin{equation}
M = \frac{3 \pi}{8} m, \qquad J_i = \frac23 M a_i = \frac{\pi}{4} m a_i, \qquad \Omega_i = \frac{a_i}{r_+^2 + a_i^2}.
\end{equation}
The temperature and the entropy are
\begin{equation}
T = \frac{r_+^4 - a_1^2 a_2^2}{2 \pi r_+ (r_+^2 + a_1^2) (r_+^2 + a_2^2)}, \qquad
S = \frac{\pi^2 (r_+^2 + a_1^2) (r_+^2 + a_2^2)}{2 G r_+}.
\end{equation}
The Smarr formula and first law in these solutions are as follows:
\begin{equation}
\frac23 M = T S + \Omega_1 J_1 + \Omega_2 J_2, \qquad
\delta M = T \delta S + \Omega_1 \delta J_1 + \Omega_2 \delta J_2.
\end{equation}

\subsection{Quantum improved black holes with different angular momenta}

For the quantum improved black holes, we assume that the Newton coupling is a function of the horizon radius
and two angular momenta, $G(r_+, a_1, a_2)$. The temperature calculated again from the surface gravity $\kappa$
in Eq.~\eqref{surfaceg} is then
\begin{equation}
\label{def:temp:MPdiff}
T = \frac{2 (r_+^4 - a_1^2 a_2^2) G(r_+, a_1, a_2) - r_+ (r_+^2 + a_1^2) (r_+^2 + a_2^2) \partial_+
 G(r_+, a_1, a_2)}{4 \pi r_+ (r_+^2 + a_1^2) (r_+^2 + a_2^2) G(r_+, a_1, a_2)}.
\end{equation}
From the first law of thermodynamics, repeating the similar steps to Eq.~\eqref{Kerrent}, we find
\begin{eqnarray}
&\!\!& \partial_+ S = \frac{\pi^2 (3 r_+^4 + a_1^2 r_+^2 + a_2^2 r_+^2 - a_1^2 a_2^2)}{2 r_+^2 G(r_+, a)},
\\
&\!\!& \partial_{a_1} S = \frac{\pi^2 (r_+^2 + a_2^2) \left[ 4 a_1 (r_+^4 - a_1^2 a_2^2) G(r_+, a_1, a_2)
- (r_+^2 + a_1^2) (3 r_+^4 + a_1^2 r_+^2 + a_2^2 r_+^2 - a_1^2 a_2^2) \partial_{a_1} G(r_+, a_1, a_2) \right]}
{2 r_+ G(r_+, a_1, a_2) \left[ 2 (r_+^4 - a_1^2 a_2^2) G(r_+, a_1, a_2) - r_+ (r_+^2 + a_1^2) (r_+^2 + a_2^2)
 \partial_+ G(r_+, a_1, a_2) \right]},
\\
&\!\!& \partial_{a_2} S = \frac{\pi^2 (r_+^2 + a_1^2) \left[ 4 a_2 (r_+^4 - a_1^2 a_2^2) G(r_+, a_1, a_2)
- (r_+^2 + a_2^2) (3 r_+^4 + a_1^2 r_+^2 + a_2^2 r_+^2 - a_1^2 a_2^2) \partial_{a_2} G(r_+, a_1, a_2) \right]}
{2 r_+ G(r_+, a_1, a_2) \left[ 2 (r_+^4 - a_1^2 a_2^2) G(r_+, a_1, a_2)
 - r_+ (r_+^2 + a_1^2) (r_+^2 + a_2^2) \partial_+ G(r_+, a_1, a_2) \right]}.~~~~
\end{eqnarray}
From each pair of $(r_+, a_1, a_2)$, we have three consistency conditions.
Unfortunately it is difficult to reexpress any condition into a total derivative as~\eqref{co_Kerr}.
But we find that if we impose the following constraints
\begin{eqnarray}
&& (3 r_+^4 + a_1^2 r_+^2 + a_2^2 r_+^2 - a_1^2 a_2^2) \partial_{a_1} G(r_+, a_1, a_2)
= 2 a_1 r_+ (r_+^2 + a_2^2) \partial_+ G(r_+, a_1, a_2),
\\
&& (3 r_+^4 + a_1^2 r_+^2 + a_2^2 r_+^2 - a_1^2 a_2^2) \partial_{a_2} G(r_+, a_1, a_2)
= 2 a_2 r_+ (r_+^2 + a_1^2) \partial_+ G(r_+, a_1, a_2),
\end{eqnarray}
$\partial_{a_1} S$ and $\partial_{a_2} S$ can be significantly simplified as
\begin{equation}
\partial_{a_1} S = \frac{\pi^2 a_1 (r_+^2 + a_2^2)}{r_+ G(r_+, a_1, a_2)}, \qquad
\partial_{a_2} S = \frac{\pi^2 a_2 (r_+^2 + a_1^2)}{r_+ G(r_+, a_1, a_2)},
\end{equation}
and all the three consistencies are satisfied.
The constraints lead to $G(r_+, a_1, a_2) = G\left( 2\pi^2 \frac{(r_+^2 + a_1^2)(r_+^2 + a_2^2)}{r_+} \right)$
and the corresponding entropy is
\begin{equation}
S = \int \frac{dx_+}{4G(x_+)}, \qquad
x_+ \equiv 2 \pi^2 \frac{(r_+^2 + a_1^2)(r_+^2 + a_2^2)}{r_+}.
\label{area:MP}
\end{equation}
Here again $x_+$ is the horizon area in 5D. Thus this is a universal formula for the entropy.

Note again that $x_+$ can be generalized to the area at a fixed radius $r$ $(a_1 a_2 = r_+ r_-$)
{\small \begin{equation}
x = \frac{4\pi^2}{3} \frac{\sqrt{(r^2 \!+\! a_1^2)^3 [r^4 \!+\! (a_1^2 \!+\! a_2^2) r^2 \!+\! 2 a_1^2 a_2^2
 \!+\! a_2^4 \!+\! a_2^2 (r_+^2 \!+\! r_-^2)]^3} \!-\! \sqrt{(r^2 \!+\! a_2^2)^3 [r^4 \!+\! (a_1^2
\!+\! a_2^2) r^2 \!+\! 2 a_1^2 a_2^2 \!+\! a_1^4 \!+\! a_1^2 (r_+^2 \!+\! r_-^2)]^3}}
{(a_1^2 - a_2^2) (r^2 - r_+^2) (r^2 - r_-^2)}.
\label{genarea:MP}
\end{equation}
}
This formula reproduces (4.15) in the limit $r\to r_+$,
and it is natural to make the identification
\bea
k^3 = \frac{\xi^3}{x},
\eea
for outside the horizon.

Here we have a similar subtlety to that for Kerr black holes.
We may first substitute $r_+$ into $G(r)$ and afterwards take $\partial_+$ when we calculate surface gravity.
We may regard this as our definition of the temperature.
Alternatively we could adopt the area obtained by replacing $r_+$ by $r$ in \eqref{area:MP},
just as for Kerr black holes.
A possible problem in this latter choice is that the area becomes imaginary for $r^2 <0$,
whereas the area~\eqref{genarea:MP} does not.
(This was not a problem for Kerr black holes because the range of $r$ is limited to $r^2 \geq 0$.)
However this may not be a problem since we expect that the behavior near the singularity may be modified.

\subsection{Quantum improved black holes with equal angular momenta}

Let us consider the special case of equal angular momenta. We set $a_1 = a_2 = a$ and then we have
$\Pi = (r^2 + a^2)^2, \; F = r^2/(r^2 + a^2)$ and the metrics are
\begin{eqnarray}
ds^2 &=& - dt^2 + \frac{\Sigma}{\Delta} dr^2 + \frac{r^2 + a^2}4 \left( \sigma_1^2
+ \sigma_2^2 + \sigma_3^2 \right) + \frac{G m}{r^2 + a^2} \left( dt - \frac{a}2 \sigma_3 \right)^2,
\nonumber\\
&& \Delta = (r^2 + a^2)^2 - G m r^2, \qquad \Sigma = r^2 (r^2 + a^2).
\end{eqnarray}
Here we have made the coordinate transformation $\theta = 2 \theta', \; \phi = \phi_2 - \phi_1, \;
\psi = \phi_2 + \phi_1$ and the range of the coordinates are $0 < \theta < \pi, \; 0 < \phi < 2 \pi, \; 0 < \psi
 < 4 \pi$. We have also defined
\begin{equation}
\sigma_1 = - \sin\psi d\theta + \cos\psi \sin\theta d\phi, \quad
\sigma_2 = \cos\psi d\theta + \sin\psi \sin\theta d\phi, \quad
\sigma_3 = d\psi + \cos\theta d\phi, \quad d\sigma_i = \frac12 \epsilon_i{}^{jk} \sigma_j \wedge \sigma_k.
\end{equation}
The Kretschman scalar is given by
\begin{equation}
\label{def:K:MP:equal}
K = R_{\alpha\beta\mu\nu} R^{\alpha\beta\mu\nu} = \frac{24 G^2 m^2 (3 r^2 -a^2)(r^2 - 3 a^2)}{(r^2 + a^2)^6}.
\end{equation}
Note that in our coordinates, the curvature singularity of the classical geometry is located at $r^2 = -a^2 < 0$.
Note also that the above Kretschman scalar is independent of the polar angle coordinate and is expressed merely
in terms of the radial coordinate $r$, in contrast to the cases of the Kerr and general Myers-Perry solutions.

Now let us consider running coupling $G \to G(r, a)$, and assume again that the temperature $T$ is given
by the surface gravity of the quantum improved metric.
We have
\begin{equation}
\label{def:temp:MPeqam}
T = \frac{2 (r_+^2 - a^2) G(r_+, a) - r_+ (r_+^2 + a^2) \partial_+ G(r_+, a)}{4 \pi r_+ (r_+^2 + a^2) G(r_+, a)},
\end{equation}
and
\begin{eqnarray}
&& \partial_+ S = \frac{\pi^2 (r_+^2 + a^2) (3 r_+^2 - a^2)}{2 r_+^2 G(r_+, a)},
\\
&& \partial_a S = \frac{\pi^2 (r_+^2 + a^2) \left[ 8 a (r_+^2 - a^2) G(r_+, a)
- (r_+^2 + a^2) (3 r_+^2 - a^2) \partial_a G(r_+, a) \right]}{2 r_+ G(r_+, a)
\left[ 2 (r_+^2 - a^2) G(r_+, a) - r_+ (r_+^2 + a^2) \partial_+ G(r_+, a) \right]}.
\end{eqnarray}
The consistency condition can be expressed as
\bea
\label{cMP}
\partial_+ \frac{r_+ (3 r_+^2 - a^2) \partial_a G(r_+, a) - 4 a r_+^2 \partial_+ G(r_+, a)}{2 (r_+^2 - a^2) G(r_+, a)
 - r_+ (r_+^2 + a^2) \partial_+ G(r_+, a)} = 0,
\eea
giving
\bea
\frac{r_+ (3 r_+^2 - a^2) \partial_a G(r_+, a) - 4 a r_+^2 \partial_+ G(r_+, a)}{2 (r_+^2 - a^2) G(r_+, a)
- r_+ (r_+^2 + a^2) \partial_+ G(r_+, a)} = f(a),
\eea
which simplifies $\partial_a S$:
\begin{equation}
\partial_a S = \frac{\pi^2 (r_+^2 + a^2) \left[ 4 a r_+ - (r_+^2 + a^2) f(a)\right]}{2 r_+^2 G(r_+, a)}.
\end{equation}
For $f(a) = 0$, the solution is $G(r_+, a) = G\left( 2\pi^2 \frac{(r_+^2 + a^2)^2}{r_+} \right)$, and
the corresponding entropy is
\begin{equation}
S = \int \frac{dx_+}{4G(x_+)}, \qquad
x_+ \equiv 2\pi^2 \frac{(r_+^2 + a^2)^2}{r_+}.
\end{equation}
Again $x_+$ is the horizon area and the consistency requires
that the Newton coupling should be a function of the area. The identification for the energy scale
$k_+ = k(x_+)$ at the horizon is
\bea
k_+^3 = \frac{\xi^3}{x_+} = \frac{\xi^3 r_+}{2\pi^2(r_+^2 + a^2)^2}.
\eea
which gives
\bea
G(r_+, a) = \frac{G_0}{1 + \omega G_0 k_+^3} = \frac{G_0 x_+}{x_+ + \tilde{\omega} G_0}
= \frac{G_0 (r_+^2 + a^2)^2}{(r_+^2 + a^2)^2 + \tilde{\omega} G_0 r_+},
\eea
with $\tilde{\omega} = \xi^3 \omega/(2\pi^2)$.
The associated entropy is
\begin{equation}
S = \frac{\pi^2 (r_+^2 + a^2)^2}{2 G_0 r_+} + \frac{\pi^2 \tilde{\omega}}{2} \ln\frac{(r_+^2 + a^2)^2}{r_+}.
\end{equation}

One can generalize the $x_+$ to area at any fixed radius $r$ as
\begin{equation}
x = 2\pi^2 \frac{\sqrt{(r^2+a^2) \left[ (r^2+a^2)^2 r_+^2 + (r_+^2+a^2)^2 a^2 \right]}}{r_+},
\end{equation}
and the identification for the energy scale $k = k(x)$ near and outside the horizon may be chosen,
on dimensional grounds, as
\bea
k^3 = \frac{\xi^3}{x} = \frac{\xi^3 r_+}{2\pi^2\sqrt{(r^2+a^2) \left[ (r^2+a^2)^2 r_+^2 + (r_+^2+a^2)^2 a^2 \right]}}.
\eea
which gives
\bea
G(r, a) = \frac{G_0}{1 + \omega G_0 k^3} = \frac{G_0 x}{x + \omega \xi^3 G_0}.
\label{4.33}
\eea
In contrast to 4D Kerr black hole, near the singularity, $r^2 \to -a^2$, the energy scale diverges
like $k^3 \to (r^2 + a^2)^{-1/2}$, but this is not fast enough to resolve singularity which needs
$k^3 \to (r^2 + a^2)^{-4}$.

An interesting choice that extrapolates the behavior \eqref{4.33} for large $x$ and
resolve singularity at $\rho^2 \equiv r^2 + a^2 \to 0$ is the identification
\bea
k^3 = \frac{\xi^3}{x} \left( 1 + \frac{G_0}{x} \right)^7 \quad \to \quad
\frac{\xi^3 r_+^8 G_0^7}{(2 \pi^2)^8 (r_+^2 + a^2)^8 a^8} \rho^{-8} \quad \textrm{as} \quad \rho^2 \to 0,
\label{scaleid:mp:equal}
\eea
which gives
\bea
G(r, a) = \frac{G_0}{1 + \omega G_0 k^3} \to \frac{(r_+^2 + a^2)^8 a^8}{\tilde{\omega} r_+^8 G_0^8} \rho^8, \qquad
\tilde{\omega} = \frac{\omega \xi^3}{(2 \pi^2)^8}.
\eea
Therefore the source term $G(r, a) \, m$, as $\rho \to 0$, approaches to zero fast enough
\begin{eqnarray}
&& G(r, a) \, m = \frac{8 G M}{3 \pi} \to \frac{8 M (r_+^2 + a^2)^8 a^8}{3 \pi \tilde{\omega} G_0^8 r_+^8} \rho^8,
 \qquad
\Delta \to \rho^4 \left( 1 + O(\rho^4) \right),
\nonumber\\
&& \Sigma = - a^2 \rho^2 \left( 1 - \frac{\rho^2}{a^2} \right), \qquad dr^2
 = \frac{\rho^2}{r^2} d\rho^2 \to - \frac{\rho^2}{a^2} d\rho^2 \left( 1 + \frac{\rho^2}{a^2} + O(\rho^4) \right).
\end{eqnarray}
Then the geometry becomes (the $O(\rho^2)$ terms in $\Sigma dr^2/\Delta$ are canceled out)
\begin{equation}
ds^2 \to - dt^2 + d\rho^2 + \frac{\rho^2}4 \left( \sigma_1^2 + \sigma_2^2 + \sigma_3^2 \right) + O(\rho^4),
\end{equation}
which is flat. The singularity appears to be resolved.

\section{Kaluza-Klein Black Strings}
\label{KKBH}

Let us consider our final example of the Kaluza-Klein black strings.
We denote the spatial dimension by $n$ with the dimension of the total spacetime being $d = n + 1$.
Our solutions are
\begin{eqnarray}
&& ds^2 = - f(r) dt^2 + \frac{dr^2}{f(r)} + r^2 d\Omega_{n-2}^2 + dz^2,
\nonumber\\
&& f(r) = 1 - \frac{G \mu}{r^{n-3}},
\end{eqnarray}
with a constant $\mu$.
This is a vacuum solution, and one can extend the $(n+1)$-th dimension to Ricci flat space of more dimensions.
However, for simplicity we restrict to the case that the extra dimension is just one-dimensional circle with
period $L$.

The mass and tension for this classical solution is given by (see, e.g.,~\cite{Kastor:2006ti, Harmark:2007md}),
\begin{equation}
M = \frac{\Omega_{n-2} L}{16 \pi} (n - 2) \mu, \qquad
\mathcal{T} = \frac{\Omega_{n-2}}{16 \pi} \mu,
\end{equation}
where $\Omega_{n-2}$ is the area of unit sphere $S^{n-2}$.
The temperature and entropy are
\begin{equation}
T = \frac1{4 \pi} \frac{d f(r)}{dr}\Big|_{r = r_+} = \frac{n - 3}{4 \pi r_+}, \qquad
S = \frac{\Omega_{n-2} L r_+^{n-2}}{4 G}.
\end{equation}
The first law and the Smarr formula are
\begin{equation}
d M = T dS + \mathcal{T} dL, \qquad
M = T S + \mathcal{T} L.
\end{equation}

The quantum improved solution is obtained by replacing the Newton coupling by the running coupling $G = G(r, L)$,
and the temperature is then given as
\begin{equation}
T = \frac{(n - 3) G(r_+, L) - r_+ \partial_+ G(r_+, L)}{4 \pi r_+ G(r_+, L)} ,
\end{equation}
and
\begin{equation}
\partial_+ S = \frac{(n - 2) \Omega_{n-2} L r_+^{n-3}}{4 G(r_+, L)}, \qquad
\partial_L S = \frac{\Omega_{n-2} r_+^{n-2} \left[ (n - 3) G(r_+, L)
 - (n - 2) L \partial_L G(r_+, L) \right]}{4 G(r_+, L) \left[ (n - 3) G(r_+, L) - r_+ \partial_+ G(r_+, L) \right]}.
\label{kkentropy}
\end{equation}
The consistency condition gives the constraint
\bea
r_+ \partial_+ G(r_+, L) = (n - 2) L \partial_L G(r_+, L),
\eea
which leads to
\bea
G(r_+, L) = G(x_+), \qquad x_+ \equiv \Omega_{n-2} \, L \, r_+^{n-2},
\eea
the latter being the horizon area!
Thus we again find that the Newton coupling should be a function of the area.
We also find the entropy is given by the universal formula
\bea
S = \int \frac{dx_+}{4G(x_+)}.
\eea
The natural identification outside the horizon is again
\bea
k^{n-1} = \frac{\xi^{n-1}}{x}, \qquad x \equiv \Omega_{n-2} \, L \, r^{n-2},
\eea
with $x$ being the area at fixed radius.
Note that the subtlety that appeared in rotating solutions does not exist for static solutions.
The running coupling relation~\eqref{Gk} in higher dimensions is
\begin{equation}
G(k) = \frac{G_0}{1 + \omega G_0 k^{n-1}},
\end{equation}
which leads to
\begin{equation}
G(r) = \frac{G_0 L r^{n-2}}{L r^{n-2} + \tilde{\omega} G_0}, \qquad
\tilde{\omega} = \frac{\omega \xi^{n-1}}{\Omega_{n-2}},
\end{equation}
and the associated entropy is
\begin{equation}
S = \frac{\Omega_{n-2} L r_+^{n-2}}{4 G_0} + \frac{\tilde{\omega} \Omega_{n-2}}4 \ln\left( L r_+^{n-2} \right).
\end{equation}
Again the first term is the standard Bekenstein-Hawking formula and the second term is
the quantum correction.

It may be interesting to notice that the FRG gives a known nonsingular black hole solution.
For the five-dimensional version ($n = 4$) with the identification ($x = 2 \pi^2 L r^2$)
\bea
k^3 = \frac{\xi^3}{x} \left( \frac{G_0}{x} \right)^{1/2} = \frac{\tilde\xi^3 G_0^{1/2}}{L^{3/2} r^3}, \qquad
\tilde{\xi} = \frac{\xi}{\sqrt2 \pi},
\eea
the quantum improved black string is given by the metric function,
\bea
\label{metric:regular}
f_q(r) = 1 - \frac{\mu G_0 L^{3/2} r^2}{L^{3/2} r^3 + \tilde \omega G_0^{3/2}}, \qquad
\tilde{\omega} = \omega \tilde{\xi}^3.
\eea
The four-dimensional spacetime is nothing but Hayward's nonsingular black hole~\cite{H}.

\section{Conclusions and discussions}
\label{DC}

In this paper, we have proposed the first law of black hole thermodynamics as the physical principle to determine
the scale identification. Our study of the consistency of the thermodynamics has revealed
that the energy scale should be related to some function of the area, at least near the horizon.
We find that this leads to a universal formula for the quantum improved black hole entropy,
which reproduces the well-known Bekenstein-Hawking formula when the Newton coupling is constant.
Interesting enough, the result is valid for all the solutions possible in four dimensions, and also
for some of higher-dimensional solutions.
This identification may be naturally extended away from the horizon toward large $r$ since the area is
a geometrical quantity and is independent of the angles, and gives results consistent for classical case
for large $r$ where we expect that the quantum effects are not important.

As for the short distances inside the horizon, we can naturally extend our proposed identification from
the horizon toward the singularity for the cases of spherically symmetric black holes and
static Kaluza-Klein black strings. In fact, for these cases, we can find the identifications that
resolve the singularity. For example, we have seen that our quantum improvement reproduces
the known nonsingular black hole model (\ref{metric:regular}) in the $4$-dimensional part of
the Kaluza-Klein black string. Furtherore, even in the rotating case, for the Myers-Perry solutions
with equal angular momenta, we have provided the concrete example (\ref{scaleid:mp:equal})
of the scale identification which is compatible with the first law of thermodynamics and
resolves the singularity. However, we have also found that our identification would need to be modified
inside the horizon for the Kerr black hole, since high-energy limit $k \to \infty$ is not attained
for any region of our spacetime.
In such cases, the singularity would not be resolved, since the Newton coupling does not become zero.
It would be important to find some guiding principle for determining the identification near the singularity.

When attempting to derive the Hawking radiation, as far as our quantum improved black hole can be viewed
as an effective classical geometry and geometrical optics for quantum fields can be applied as in the Hawking's
original derivation~\cite{Haw}, one can naturally expect that the corresponding Hawking temperature
should be proportional to the surface gravity of the background black hole horizon.
The surface gravity $\kappa$ measures the difference between the Killing and affine parameters
along the null generators of the event horizon, and can be computed by using the formula
$\kappa^2=-\lim_{r\rightarrow r_+}(\nabla_\mu \chi_\nu)(\nabla^\mu \chi^\nu)/2$ with
$\chi^\mu$ being the horizon Killing field.
Our definition of the Hawking temperature is based essentially upon the surface gravity, but as briefly
discussed below Eq.~\eqref{area at r}, there is a subtle issue for the Kerr and other rotating black hole cases
if we use the naive extension of the area at fixed radius $r$:
The above formula for the surface gravity $\kappa$ for a quantum improved black hole involves the derivative term
$\partial_+ G(r_+)$ and when evaluating it using the area of fixed radius $r$,
we first substitute the horizon radius $r_+$ of the quantum improved geometry into $G(r)$ and
afterwards take the derivative $\partial_+$, rather than the other way around.
This procedure appears to be necessary in order to make our temperature to be consistent with the first law
of thermodynamics when the black hole is rotating.
The necessity of this procedure for the rotating case may possibly be related to the issue of whether
(and how) we can uniquely extend our area function $x_+$ and the running coupling $G(x_+)$
on the horizon to any fixed radius $r$ and $G(x)$ away from the horizon.
This may give rise to a slight difference from the surface gravity computed in the standard manner,
and in this sense, our temperature can be called a modified Hawking temperature.
We would, however, like to emphasize that our modification of the temperature is different in nature from
the necessity of modifications of temperature and entropy discussed in \cite{RT:2010}.
The latter was motivated by the attempt to remove, in an ad hoc way, some unwanted angular-coordinate-dependence
in the temperature and entropy, while our modified Hawking temperature is based on the geometric aspect and,
together with our proposed running Newton coupling, guarantees the first law for black hole thermodynamics to hold.

Alternatively we may adopt the area obtained just by replacing $r_+$ by $r$ in the horizon area.
This is conceptually more natural and simpler without using special calculational procedure.
As stated above, there is still ambiguity in the extension away from the horizon, and we need more physical
guiding principle to determine it.

Finally, in relation to the singularity resolution inside the horizon, we should note that there are different
approaches for quantum improvements, which in principle make different predictions
(see, e.g.,~\cite{BR:2000,PS:2018,Plat2020,IOY:2021,RW04}).
In the present paper, we have performed the quantum improvement at the level of solution, that is,
we first take a classical solution and then replace the Newton coupling involved in the solution with
the running one $G(x)$. This approach is simple enough but may be applicable only when the improved geometry
is not drastically different from the classical counterpart. When we are concerned with extremely strong
gravity regime, such as near the singularity, it may be more appropriate to perform quantum improvement
at the level of action, which is more natural and expected to have higher predictive power,
compared to the present approach of the improvement at the solution level.
Some related discussions have been given in Refs.~\cite{BR:2000,Plat2020}.
It remains to be seen how the idea could be realized more concretely.

\acknowledgments

The work of C.M.C. was supported by the National Science Council of the R.O.C. (Taiwan) under the grant
MOST 110-2112-M-008-009.
The work of A.I. was supported in part by JSPS KAKENHI Grants No. 21H05182, 21H05186, 20K03938, 20K03975, 17K05451,
and 15K05092.
The work of N.O. was supported in part by the Grant-in-Aid for Scientific Research Fund of the JSPS (C) No. 16K05331,
20K03980, and Taiwan MOST 110-2811-M-008-526.

\end{document}